\documentclass[conference]{IEEEtran}
\IEEEoverridecommandlockouts
\usepackage{cite}
\usepackage{amsmath,amssymb,amsfonts}
\usepackage{algorithmic}
\usepackage{graphicx}
\usepackage{textcomp}
\usepackage{xcolor}
\usepackage{booktabs}
\usepackage{subcaption}

\usepackage[breaklinks,colorlinks,bookmarks=false]{hyperref}
\usepackage{cleveref}
\usepackage[numbers]{natbib}

\usepackage{url}
\hypersetup{citecolor=blue,linkcolor=blue,urlcolor=black}

\usepackage{flushend}

\def\BibTeX{{\rm B\kern-.05em{\sc i\kern-.025em b}\kern-.08em
    T\kern-.1667em\lower.7ex\hbox{E}\kern-.125emX}}

\begin{document}

\title{COSEA: Convolutional Code Search with Layer-wise Attention}


\author{\textbf{Hao Wang},\textsuperscript{1}
\textbf{Jia Zhang},\textsuperscript{2}
\textbf{Yingce Xia},\textsuperscript{2}
\textbf{Jiang Bian},\textsuperscript{2}
\textbf{Chao Zhang},\textsuperscript{1}
\textbf{Tie-Yan Liu} \textsuperscript{2} \\
\textsuperscript{1}{Tsinghua University} \textsuperscript{2}{Microsoft Research}\\
\textsuperscript{1}hao-wang20@mails.tsinghua.edu.cn, \textsuperscript{1}chaoz@tsinghua.edu.cn, \\
\textsuperscript{2}\{zhangjia, yinxia, Jiang.Bian, tyliu\}@microsoft.com}

\maketitle


\begin{abstract}
Semantic code search, which aims to retrieve code snippets relevant to a given natural language query, has attracted many research efforts with the purpose of accelerating software development. The huge amount of online publicly available code repositories has prompted the employment of deep learning techniques to build state-of-the-art code search models. Particularly, they leverage deep neural networks to embed codes and queries into a unified semantic vector space, 
and then use the similarity between code's and query's vectors to approximate the semantic correlation between code and the query. However, most existing studies overlook the code’s intrinsic structural logic, which indeed contains a wealth of semantic information,
and fails to capture intrinsic features of codes. 

In this paper, we propose a new deep learning architecture, COSEA, which leverages convolutional neural networks with layer-wise attention to capture the valuable code’s intrinsic structural logic. To further increase the learning efficiency of COSEA, we propose a variant of contrastive loss for training the code search model, where the ground-truth code should be distinguished from the most similar negative sample. 
We have implemented a prototype of COSEA.
Extensive experiments over existing public datasets of Python and SQL have demonstrated that COSEA can achieve significant improvements over state-of-the-art methods on code search tasks.
\end{abstract}


\section{Introduction}\label{section1}
Continuously expanding scale and soaring complexity of modern software projects have been making it increasingly difficult for efficient development and subsequent maintenance. In reality, experienced software engineers prefer to reuse stable and well-tested codes to avoid reinventing the wheel. 
In order to find desired codes fulfilling the certain program functionality, many software engineers mainly rely on search engines and/or community Q\&A sites, by using which they attempt to manually locate the relevant codes from retrieved web pages. This approach is undoubtedly inefficient. 
Fortunately, source code management communities, GitHub as the most famous one, have emerged as a popular and effective platform for a wide range of code retrieval. The importance of such community sites is magnified as they host a huge amount of high-quality code repositories~\citep{Cambronero2019,Yao2019,Ye2020}. 

However, to make this immense body of code repositories accessible, it remains challenging to efficiently and effectively retrieve archived codes relevant to natural language queries, mainly due to the semantic gap between the source codes’ well-formatted logic flow and concise natural language queries.  
Inspired by multi-modal embedding \citep{karpathy2015deep, Xu2015, frome2013devise}, previous work~\citep{Sachdev2018,bojanowski2017enriching,gu2018deep} introduced deep learning methods to solve the code search task. Specifically, they propose to jointly learn two neural networks that can embed codes and queries into a unified semantic vector space, in which they can calculate the correlation between codes and queries based on the vector similarity. 
While proving the effectiveness of applying deep learning techniques into code search, most of them simply treat both codes and queries the same as plain text and inappropriately overlook the code's intrinsic structural logic which in fact contains invaluable information about the code’s functionality and semantics. This indeed limits the capacity of the deep learning techniques over code search.

To better address the semantic code search problem, 
we believe there are three research questions that need to be answered.
\paragraph{Research Question 1 (RQ1)}
\textit{How to capture the semantic information of code more accurately?}

Representations of code and query are required for the state-of-art neural code search framework. Compared with natural language queries, the structure and logic of code is much more complex. At present, the existing natural language processing techniques can extract the semantic information of queries well. However, there is no unique answer to the deep learning model for code representation. How to represent the code with complex structures well is the most critical challenge in the code search task. We believe that a better code semantic information extraction will lead to great improvement in the code search task. In section \ref{rq1}, the evaluation results show that well-designed network 
indeed can improve the ability of code representation.



\paragraph{Research Question 2  (RQ2)}
\textit{How to improve the training speed of the code search task?}

The prevalent approach to code search learning maps a code snippet or a natural language query to a vector representation via recurrent neural networks. As we all know, the weaknesses of RNN-based models is that the computing process can not be parallel, which is unable to make full use of GPU resources and greatly limits the speed of large-scale training. Based on the above discussion, adopting a model which can make computations fully parallelized is a solution to faster training. 
The evaluation results in Section \ref{rq2}  confirm that a parallel model can be applied to code search and can achieve better performance. 

Moreover, the learning object (e.g., loss functions) of supervised code search also has a big impact on the training speed of the model. 
The evaluation results in Section \ref{rq2} also prove that a better learning object can lead to faster convergence and better performance.

\paragraph{Research Question 3 (RQ3)}
\textit{What does the deep learning model actually learn when dealing with the code search task?}

Deep learning models have the common problem of poor interpretability. Most of the previous works do some quantitative analysis through the examples of results. However, they are unable to describe the internal state of the model or to explain the representation of code snippets in detail. In order to prove that the code search model really learns how to represent the semantic information of code and natural language query, rather than just memorizing the association on the dataset, it is necessary to do some analysis on the internal representation of the input code and query in the code search model. Our case study in Section \ref{rq3} gives examples of how the deep learning model represents code and queries.






\vspace{0.1cm}
{\bf Our solution:}
In this paper, we present a novel deep learning architecture, COSEA, to address the  semantic code search problem.
It leverages convolutional neural networks with layer-wise attention to capture the  valuable code’s intrinsic structural logic. 

In order to enhance deep learning models to better leverage the code's structural information, we turn to the intuition that how experienced human programmers perceive code snippets. 
To understand the functionality of code snippets, experienced programmers never recklessly scan each token one by one. Instead, they are inclined to follow program grammar to divide code snippets into a sequence of code blocks, which has more logical integrity. Sometimes, programmers can even understand the whole meaning of a code snippet without reading all tokens, as one special code block could be enough to tell the functionality of the code snippet. That is, code blocks are crucial to understanding the comprehensive meaning of a code snippet.
In other words, code snippets' semantic information is composed of the combination of code blocks. However, previous works just ignored this important characteristic of codes and treated them as flat sequences.
To answer the first research question RQ1, we propose a well-designed network to capture features of code blocks and combine them together.

To answer the second research question RQ2, we adopt a convolutional neural network (CNN) to learn the block representation for code search, since CNN can efficiently capture the locality information in code snippets. Each convolutional layer can aggregate local information based on shorter block representation output by previous layers. After the aggregation process, we introduce the layer-wise attention to learn the respective weight on each output representation vector and re-scale this vector with this weight before being fed into the next layer. These weights are used to let the successive convolutional layers pay attention to more important representation vectors when composing longer block representations for code. 
Moreover, we propose a variant of contrastive loss,
which distinguishes the  ground-truth  code  from  the most  similar  negative  sample,
to further increase the learning efficiency and performance of the code search task in the training process. 

To answer the third research question RQ3, we examine the internal of the model.
COSEA  explicitly  contains  the  layer-wise attention for learning relationship among code blocks and the attention for learning relationship among words of queries.
We thus check  the  attention  representation  of  COSEA to qualitatively analyze what COSEA has learned.

We trained a model of COSEA based on the high-quality dataset StaQC\cite{yao2018staqc}, which is the largest high-quality dataset in SQL and Python domain obtained from Stack Overflow by using data mining techniques.
We then compare our model with state-of-the-art models, including CoaCor\citet{Yao2019}, CO3~\citet{Ye2020}, UNIF~\citet{Cambronero2019}, and CODEnn~\citet{gu2018deep}.
The evaluation  results showed that,  
COSEA has a significant advantage over other baseline methods on StaQC-Python and StaQC-SQL under all evaluation metrics.

In summary, we make the following contributions in this paper:
\begin{itemize}
\item We propose a novel model COSEA to learn the relationship between code snippets and natural language queries. To our best knowledge, COSEA is the first to leverage a deep convolutional neural network with layer-wise attention to learn code representations used in code search tasks. Compared with existing RNN-based models, COSEA yields faster convergence.
\item  We propose a min-max contrastive learning objective for training code search models, which can result in faster convergence and better performance.
\item  We evaluate our approach on two code search datasets StaQC-Python and StaQC-SQL collected from StackOverflow by previous work~\citep{yao2018staqc}. COSEA can outperform state-of-the-art methods on both datasets.
\item We conduct a comprehensive ablation study to show the effects of layer-wise attention and min-max contrastive learning objective, respectively, and we further examine how COSEA captures code block information through some case studies with visualization.
\end{itemize}

\section{Background}
We adopt some advanced techniques from deep learning and natural language process~\citep{sutskever2014sequence, gehring2017convolutional, Mikolov2013, bahdanau2014neural, vaswani2017attention}, we first discuss the background of these techniques in this section.
\subsection{Token Embedding} 
Embedding~\citep{turian2010word, Mikolov2013} is a crucial technique in deep learning, which can be used for learning vector representations of entities like words, images and videos. It is beneficial as similar entities have vectors close to each other. 

\begin{figure}[htbp]
 \includegraphics[width=0.9\linewidth]{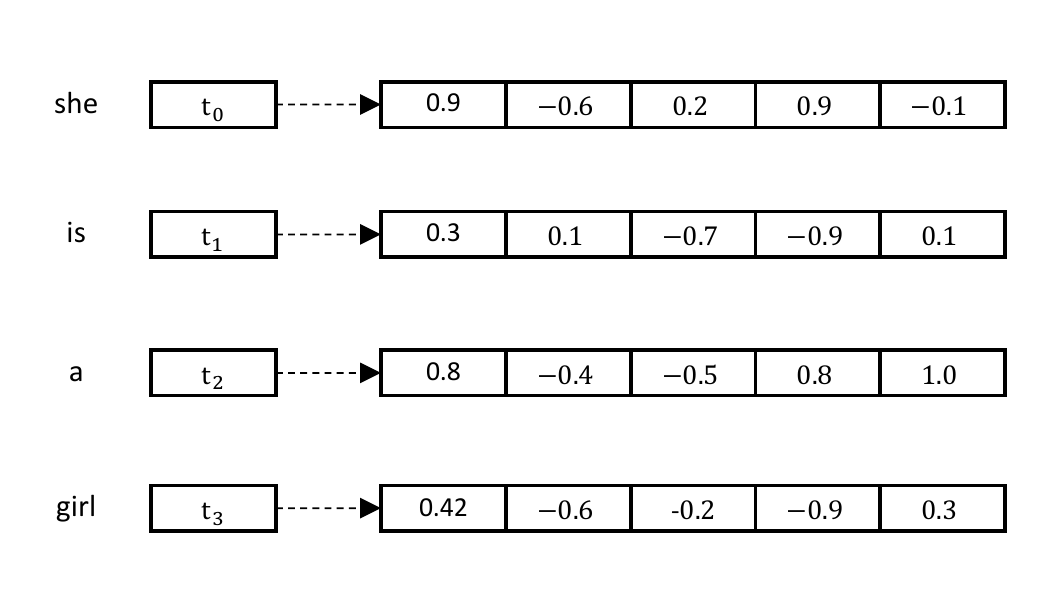}
 \caption{An example of Word Embedding.}\label{fig:embedding}
\end{figure}

Word embedding is a typical embedding technique, which converts a word into a low-dimensional semantic vector. Figure \ref{fig:embedding} gives an example of Word Embedding. Denote a sentence as $w_{1:n}=[w_1, w_2, ... w_n]$, while $n$ is the sentence length. The sentence will be converted into a sequence of word vectors $[e_1, e_2, ... e_n]$ via a word embedding look-up table $W_e \in \mathbb{R}^{V\times D}$. After that, we can calculate the semantic relation between two words using the word embedding vectors. Likewise, we can use the word embedding vectors to generate embedding for a sequence. A straightforward way is to view it as a bag of words and sum up all its word vectors\citep{le2014distributed}.

Similar to word embedding, we can extract tokens from code snippets and produce a token dict. We can obtain the token embedding of the code with the same procedure as word embedding.


\subsection{CNN for Sequence Representation}

The Convolutional Neural Networks(CNN)\citep{krizhevsky2012imagenet} is a prevalent deep neural network proposed in computer vision. \citet{kim2014convolutional} first introduced CNN to build representations for sentences. 
After obtaining the word embedding vectors of each word in a sentence, traditional methods usually use the bag-of-words~\citep{le2014distributed} (BOW) technique to represent the sentence. However, BOW-based methods ignore the order of the words in sentences. The CNN-based method takes word orders into consideration, and it can capture the N-Gram information of the sentence.

Given a sequence of word vectors $[e_1, e_2, ... e_n]$, denote the contextual word representation of the $i$-th word is $c_i$, which is calculated by:
\begin{equation}
\begin{aligned}
    c_i = f(W \ast e_{i-K:i+K} + b),
\end{aligned}
\end{equation}
where $e_{i-K:i+K}$ is the vectors of the $i-k$-th to the $i+K$-th words, $f$ is the activation function, $\ast$ is the convolution operator, $K$ is the window size of filter, $W$ and $b$ are the convolution kernel and bias parameter of the CNN. In order to keep the sequence length fixed, we often pad the CNN input with zero padding.\citep{gehring2017convolutional} 

After that, a sentence representation can be obtained from pooling techniques such as mean pooling.\citep{kim2017convolutional}. Mean pooling shares the same idea as BOW models that add up all the contextual word representation after convolution to represent the whole sentence. So it has the same problem that mean-pooling overlook the order of the words in a sequence. \citet{er2016attention} introduced attentive pooling for CNN. The attention mechanism is one of the most powerful concepts in natural language processing recently, is motivated by how people pay attention to words in a sentence. For example, in the sentence `COSEA: Convolutional Code Search with Layer-wise Attention', 'Attention' is much more informative than `with'. 

Unlike mean-pooling which treat every word in a sentence equally, attentive pooling will give them different weights. 
Denote the contextual word representations after convolution as $[c_1, c_2, ... c_n]$, and the weight of the $i$-th word is $\alpha_i$, which can be computed by:

\begin{equation}
\begin{aligned}
    \alpha_{i} = \frac{\exp\left(q^{\mathsf{T}} c_i\right)}{\sum^{n}_{j=1} \exp\left(q^{{\mathsf{T}}} c_j\right)},
\end{aligned}
\end{equation}
where q is a trainable query vector of the attention layer, $c_i$ is the contextual word representation of the $i$-th word. The final sentence representation $e_{s}$ is formulated as the weighted summation of the contextual word representations
\begin{equation}
\begin{aligned}
    e_{s} = \sum^n_{j=1}\alpha_{j}c_{j}
\end{aligned}
\end{equation}

\begin{figure*}[h!]
 \centering
 \includegraphics[height=0.3\textwidth]{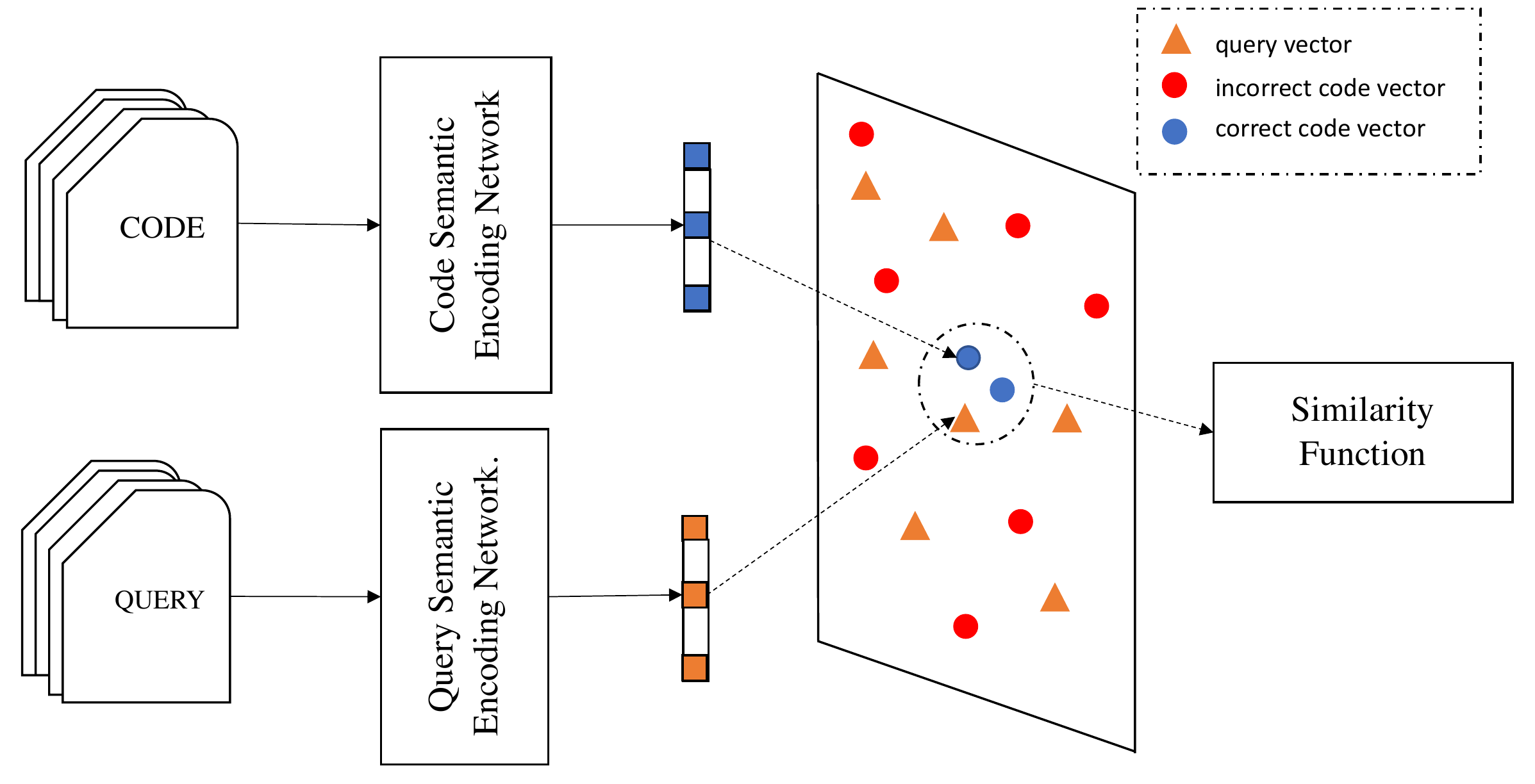}
 \caption{The overview of code search algorithm.}
\label{fig:overview}
\end{figure*}

\section{Related Work}
Code search is essentially a learning-to-rank problem~\citep{liu2009learning} targeting at retrieving code with the coming natural language queries. There have been multiple studies borrowed ideas from information retrieval systems and deep learning for code search tasks. Most of the IR-based approaches focus on query expansion~\citep{efthimiadis1996query} and query reformulation, and treat code snippets as a collection of documents. For example,  \citet{mcmillan2011portfolio} proposed Portfolio, which uses PageRank in Function Call Graph (FCG) of code snippets to build representation for code functions. \citet{Haiduc2013} trained a model to recommend the reformulation strategy based on the query property and reformulated the query to get better performance. \citet{lu2015query} proposed to employ query expansion for code search with the synonyms generated from WordNet~\citep{miller1995wordnet}. \citet{lv2015codehow} introduced CodeHow, which expands the query with the APIs and applies the Extended Boolean model to search codes. Overall, these methods mainly rely on syntax-based keyword matching and lack understanding of the semantic information behind codes and queries. 

As introduced in Section \ref{section1}, many researchers utilize deep learning techniques on code search tasks and proposed many neural code search models~\citep{Cambronero2019}. For example, \citet{Sachdev2018} proposed NCS, an unsupervised model for code search which uses the only word embedding tool, FastText~\citep{bojanowski2017enriching}, to learn the code and query representation. \citet{gu2018deep} first introduced a supervised model CODEnn, which uses the sequence-to-sequence-based recurrent networks to embed codes and queries into the unified semantic space. CODEnn is shown to outperform other state-of-art IR-based code search models on the Java datasets collected from GitHub. \citet{Cambronero2019} argued that simple models can achieve better performance than complicated models, they invented a supervised extension for the base NCS and used a bag-of-words-based network to build a new model UNIF, which has lower complexity than CODEnn. 
Most recently, \citet{Feng2020} pre-trained a large model CodeBERT, a Transformer-based neural architecture following BERT~\citep{devlin2018bert} and RoBERTa~\citep{liu2019roberta}, and finetuned it for code search task. 
Different from CodeBERT pursuing a large-scale pre-trained model, this paper focuses on building a light-weight model by explicitly learning block representations. Moreover, the pre-trained model by CodeBERT can serve as the initialization of COSEA. 
Note that, due to the limited publicly available details (e.g. source codes, datasets, hyper-parameter settings) about CodeBERT, we take the comparison with CodeBERT and further investigation of integration with it as our future plans.

Different from the aforementioned approaches, \citet{Yao2019,Ye2020} argued that combining the multi-task learning with state-of-art code search model could improve the performance. Code annotation aims to annotate a code snippet with a natural language description. Based on CODEnn~\citep{gu2018deep}, \citet{Yao2019} proposed an improved model CoaCor, trained to generate a natural language annotation which represents the semantics of a given code snippet. CoaCor used generated annotation to improve the code search model with reinforcement learning. Inspired by CoaCor's success, \citet{Ye2020} argued that incorporating code annotation and code search can improve the performance, therefore the code generation task would also be valuable for consideration. \citet{Ye2020} proposed a new model CO3 incorporating three code tasks: code search, code annotation and code generation. Besides, CO3 used the dual learning method for code generation and code annotation as the two tasks share the duality by nature. 

\section{Our Method}\label{method}
In the code search task, given a natural language query $q$ in the query set $\mathcal{Q}$, we need to output the most related code snippets $c_1,\cdots, c_k$ from the codes set $\mathcal{C}$.
As widely adopted by previous works, to determine the relevance of a pair of query and code $(q, c)$, COSEA encoders $q$ and $c$ into a uniform semantic space through the query semantic encoding network $f_q$ and the code semantic encoding network $f_c$. Then, a similarity function $\delta(\cdot,\cdot)$ is used to calculate the correlation between $q$ and $c$. The whole framework is shown in Figure \ref{fig:overview}. When searching corresponding codes for a query, codes with higher similarity value in $\mathcal{C}$ will be output to user.

\begin{figure}[htbp]
 \includegraphics[width=0.9\linewidth]{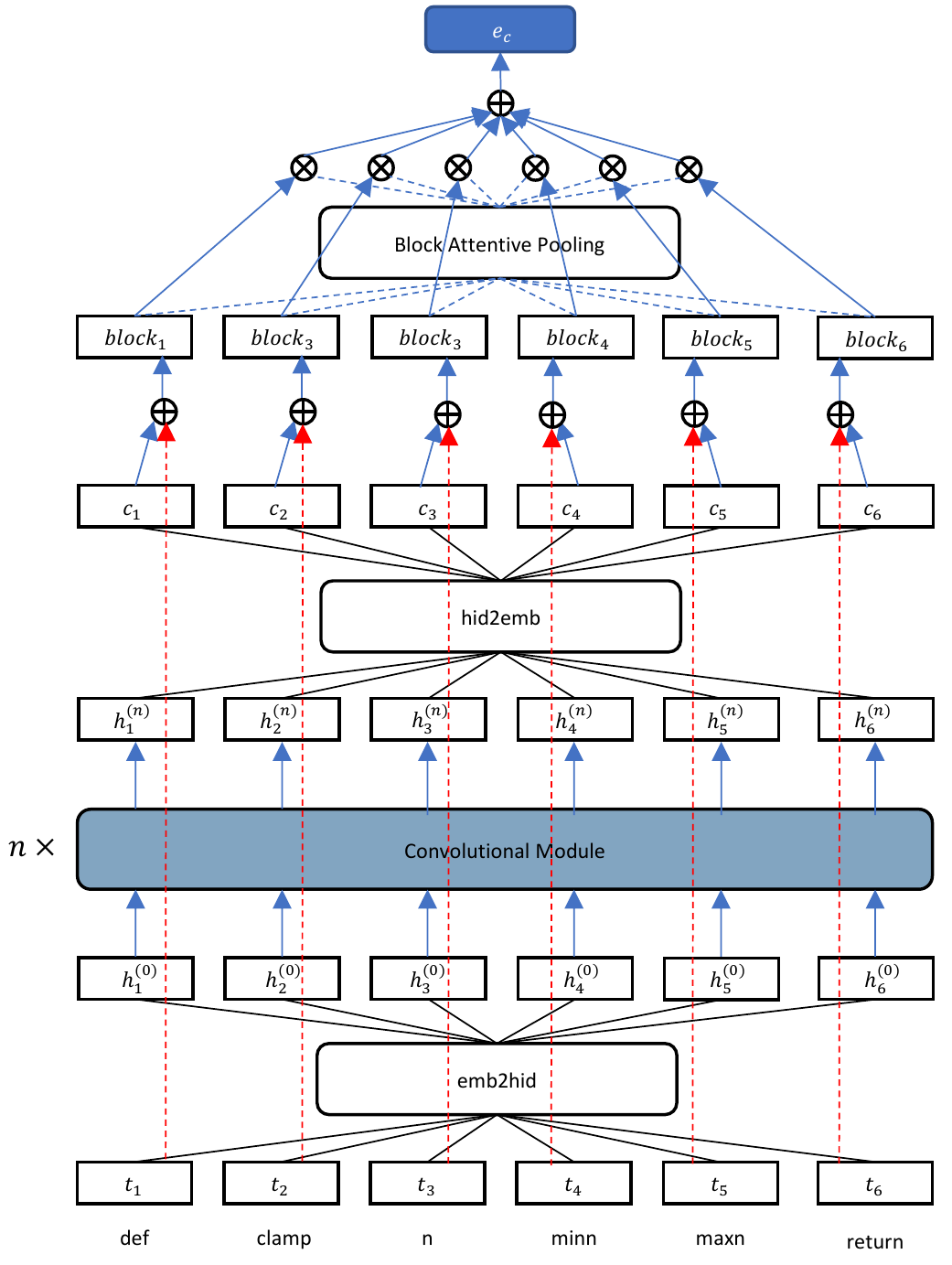}
 \caption{The architecture of code semantic encoding network.}\label{fig:COSEA}
\end{figure}

\paragraph{Code semantic encoding network}\label{encoder} In the code search task, the design of code semantic encoding network $f_c$ is the most difficult part, because the code snippets can be much longer and the semantics of them are more fuzzy. We adopt the convolutional neural network with layer-wise attention to learn the code semantic representation, whose overall architecture is shown in Figure \ref{fig:COSEA}. Generally, $f_c$ takes the embedding vectors of a code snippet $c$ as input. Through convolutional modules, we get longer and longer code block representation. 

\begin{figure}[htbp]
 \centering
 \includegraphics[width=1\linewidth]{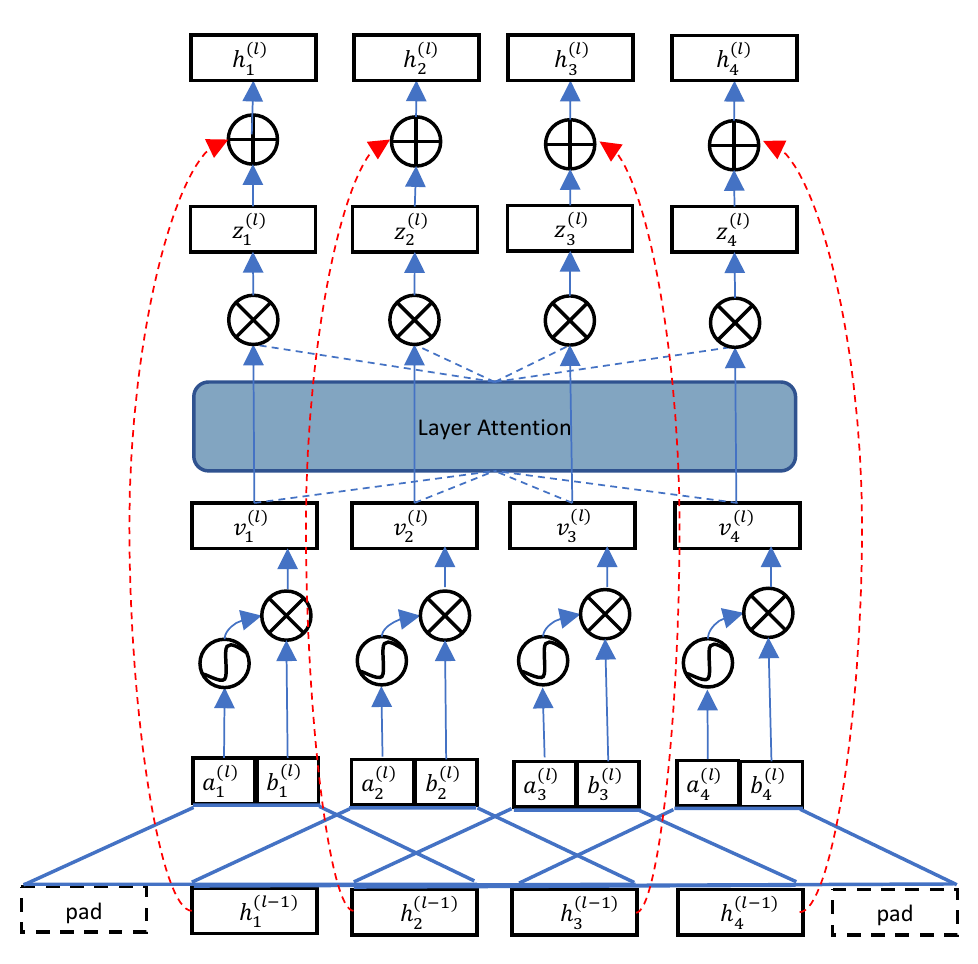}
 \caption{Detailed architecture of convolutional module with layer-wise attention.}\label{fig:conv}
\end{figure}

Finally, we perform attentive pooling on these block representation and obtain the ultimate semantic encoding $e_c$. In particular, as shown in Figure \ref{fig:conv}, each convolutional module performs convolution operation on previous output vectors with gated non-linear units as activators. Before fed into the next layer, aggregated vectors are re-scaled with learning attention weights for further representation learning.

More specifically, given a code segment $c = \{c_{1},\cdots, c_{p}\}$, $c_i$ is the $i$-th token of $c$. All the tokens will be converted into embedding vectors $\{t_1,\cdots, t_p\}$ before fed into the network. In this step, some pretrain methods can be leveraged. Similar to ConvS2S~\citep{gehring2017convolutional}, these embedding vectors are transformed into a larger space with $d$-dimension in the first layer. After that, the code semantic encoding network aggregates the local information through multiple delicate convolutional modules with layer-wise attention. This convolutional module is shown in Figure \ref{fig:conv}. In layer $l$, there is only one convolution kernel with size $k^{(l)}\times d$ and $2d$ output channels. 

Thus, the dimension of the output vector of the convolution kernel is exactly twice the dimension of the input vector. We divide the $i$-th output vector into two $d$-dimension vectors $a_i^{(l)}$ and $b_i^{(l)}$. Then we take the gated non-linear unit as the non-linear activator $\alpha(\cdot)$ to deal with each output vector as in work \citep{dauphin2017language}:
\begin{equation}
    \alpha\left(a_i^{(l)} \big\Vert b_i^{(l)}\right)= a_i^{(l)} \otimes \sigma\left(b_i^{(l)}\right),
\end{equation}
where $\otimes$ is an element-by-element multiplication and $\sigma(\cdot)$ is the gated function. Denote $\alpha\left(a_i^{(l)} \big\Vert b_i^{(l)}\right)$ as $v_i^{(l)}$ for convenience. For each vector $v_i^{(l)}$, it is an aggregated representation for $k^{(l)}$ lower-level code representations output by $(l-1)$-th layer. Before fed into next layer for longer code representation learning, we rescale $v_i^{(l)}$ with a learnable attention weight. That is 
\begin{equation}
\begin{aligned}
    \alpha_i^{(l)} = \frac{\exp\left(a^{(l){\mathsf{T}}} v_i^{(l)}\right)}{\sum_{j=1}^p\exp\left(a^{(l){\mathsf{T}}} v_j^{(l)}\right)} \text{ and }
    z_i^{(l)} = \alpha_i^{(l)}\cdot v_i^{(l)},
\end{aligned}
\end{equation}

where $a^{(l)}$ is the learnable parameters of the attention weigh in $l$-th layer. This attention weight helps higher layer to attend to important lower-level representation, which can significantly improve the performance. 

For each convolutional layer, a residual connection \citep{he2016deep} are added between the input and output to improve training efficiency and performance, i.e., the final output of the $l$-th convolutional module is
\begin{equation}
    h^{(l)}_{i} = \textsc{Conv}\left(h^{(l-1)}_{\frac{i-k}{2}},\cdots,h^{(l-1)}_{\frac{i+k}{2}}\right) + h^{(l-1)}_{i}.
\end{equation}

The output vector $h_i^{(n)}$ of the last convolutional layer is transformed back to the embedding space for residual connection with the input embedding vectors. Thus, we have $block_i = \text{hid2emb}\left(h_i^{(n)}\right)+t_i$.  Indeed, $block_i$ is the representation learnt for a code block. Finally, we can obtain the code semantic encoding $e_c$ with attentive pooling on $\{block_1, block_2, \cdots, block_p\}$.

We chose the convolutional neural networks(CNN) for two reasons. On the one hand, compared to recurrent models, CNN has the natural ability to capture locality information which can be used for capturing code blocks' information. Following the intuition that code blocks are crucial to understanding the comprehensive meaning of code snippets, we add the layer-wise attention mechanism to learn the relationship among each level code block. According to the design above, we believe the elaborately constructed code semantic embedding network can capture the semantic information of code more accurately. On the other hand, computations over CNN can be fully parallelized. This feature makes CNN-based code search models cost less time on training than RNN-based models. 

\paragraph{Query semantic encoding network.} In the query semantic encoding part, we also use the attentive pooling to generate query embedding. Given a natural language query $q = \{q_ {1}, ..., q_ {m} \}$. We can embed them into word vectors $\{o_1,\cdots,o_m\}$ with pre-training method. The attention weight $\alpha_{i}$ is calculated as follows:
\begin{equation}
    \alpha_{i} = \frac{\exp\left(a_{q}^{\mathsf{T}} W_q o_i\right)}{\sum^{m}_{j=1} \exp\left(a_{q}^{{\mathsf{T}}} W_q o_j\right)},
\end{equation}
where $a_q$ is the learnable attention weight parameters and $W_q$ is a matrix to transform the embedding vectors to a larger hidden space.

We chose the attentive pooling model to produce query vectors rather than more sophisticated networks for two reasons. Firstly, most queries are short sequences. There is little improvement to replace attentive pooling with a more sophisticated model for representing short natural language sequences. Secondly, the main cost in inference comes from query embedding because the time to embed code snippets can be done offline. For this reason, we prefer to chose attentive pooling model to search faster.  

\begin{figure*}[t]
\begin{equation}\label{eq5}
        \min_{\theta} \mathcal{L}(\theta) = \sum_{\langle q,c^+ , c^-\rangle \in \mathcal{D}}  \max \left(0, \varepsilon - \cos\left(e_q, e_{c^{+}}\right) + \cos\left(e_q, e_{c^{-}}\right)\right).
\end{equation}
\end{figure*}

\begin{figure*}[t]
\begin{equation}\label{eq6}
        \min_{\theta} \mathcal{L}_{\max}(\theta) = \sum_{{\langle q, c^+ \rangle \in \mathcal{G}}}  \max \left(0, \varepsilon - \cos\left(e_q, e_{c^{+}}\right) + \max_{c^- \in \mathcal{C}\setminus \{c^+\}} \cos\left(e_q, e_{c^{-}}\right)\right).
\end{equation}
\end{figure*}



\paragraph{Min-max contrastive learning objective.} The code search task is in fact a learning-to-rank task. We want to maximize the similarity between related query and code pairs, while minimize the similarity for unrelated pairs. Recall that $\mathcal{Q}$ and $\mathcal{C}$ as the query set and code set respectively. For any query $q\in \mathcal{Q}$, $c^+\in \mathcal{C}$ is the ground-truth code and $c^-\in \mathcal{C}$ is an arbitrary unrelated code. We define $\mathcal{G}$ as the set of all ground-truth query and code pairs, while $\mathcal{D}$ as the set composed of all generated triples $\left\langle q, c^+, c^-\right\rangle$. Then, we can define the contrastive loss on the triple $\left\langle q, c^+, c^-\right\rangle$ with Cosine similarity, i.e., $\cos\left(e_q,e_{c^+}\right) - \cos\left(e_q, e_{c^+}\right)$. Previous works \citep{Yao2019,Ye2020} adopt the learning objective as equation \ref{eq5}, where $\theta$ represents the parameters of the model, $e_q = f_e(q;\theta)$, $e_c = f_c(c;\theta)$ 
and $\varepsilon$ is a hyper-parameter to adjust the separation of similarities, which indeed affect the difficulty of the learning task. In previous works, $c^-$ is uniformly sampled from the set $\mathcal{C}\setminus \left\{ c^+\right\}$, which leads to poor data efficiency and inadequate learning. In this paper, we enhance the loss function by minimizing the largest similarity for all codes in $\mathcal{C}\setminus \left\{ c^+\right\}$, i.e., we define our learning objective as equation \ref{eq6}. This is the formal definition of our min-max contrastive learning objective. Experiments show that taking $\mathcal{L}_{\max}(\theta)$ as loss function achieves much better convergence speed and test performance. 

\section{Evaluation Methodology}

\subsection{Datasets}
\begin{table}[htbp]
\centering
\caption{Statistics of StaQC-SQL and StaQC-Python}
{\small
\label{tab:statistics}
\begin{tabular}{|c|c|c|c|c|}
\hline
 & \multicolumn{2}{c|}{StaQC-Python} & \multicolumn{2}{c|}{StaQC-SQL} \\ \hline
\#Tokens &    Code       &      Query     &      Code     &     Query      \\ \hline
1$-$20 &      14.06\%     &      99.39\%   &      17.59\%  &     98.43\%      \\ \hline
21$-$120 &    65.08\%     &      0.61\%    &      72.66\%  &     1.57\%      \\ \hline
121$-$200 &   12.07\%     &        0\%     &      6.66\%   &      0\%     \\ \hline
 $\ge$ 200 &       8.89\%      &        0\%     &      3.09\%   &      0\%     \\ \hline
\end{tabular}
}
\end{table}

We compare our work with state-of-the-art models on the high-quality dataset StaQC\cite{yao2018staqc}, which is the largest high-quality dataset in SQL and Python domain. This dataset is obtained from Stack Overflow by using data mining techniques. Table~\ref{tab:statistics} summarizes the token statistics of StaQC-Python and StaQC-SQL, each of which contains 147,546 and 119,519 pairs of question titles and code with the programming language Python and SQL, respectively. 

We use the question title submitted by users on Stack Overflow as the query text of the code snippets in the code search task. The previous works have not published the well-divided training, validation and testing datasets, so we follow the same data division in \citep{Yao2019,Ye2020}, using 75\% of the entire dataset as the training set, 10\% of it as the validation set, and the remaining 15\% as the testing set. For fair comparison, we verify the variance of the model performance through repeated experiments with random data division.

\subsection{Baseline Methods} 
We compare COSEA with state-of-the-art models on code search tasks, including:

\paragraph{CODEnn} It is proposed by \citet{gu2018deep} and considers code as a combination of function names, API sequences and other tokens. It uses bidirectional LSTM as the encoder for function names, API sequences and query tokens, and applies MLP to embed other code tokens. It uses max-pooling to get the query vector and combines all of the code snippets' features into a fully connected layer for code vector. We follow the work \citep{Yao2019} and slightly modify the architecture to adapt to our dataset.

\paragraph{UNIF} It uses FastText~\citep{bojanowski2017enriching} to pre-train code and query with training corpus, and applies the pre-trained embedding matrix to initialize the code and query embedding. After that, UNIF combines the query token embeddings with a simple average and uses a learnable attention vector to aggregate the code token embeddings. 

\paragraph{Self Attention (SA)} In order to make a comprehensive evaluation, we directly replace the bidirectional LSTM in CODEnn with the Transformer proposed in \citet{vaswani2017attention}. As this architecture is based on multi-head self-attention, we name it as SA. 

\paragraph{CoaCor} It is proposed to improve the code search model with code annotation. CoaCor first trains a code search model and a code generation model using sequence to sequence architecture base on bidirectional LSTM. Besides, it uses the pre-trained code search model as the indicator of the annotation task and trains a better code annotation model using reinforcement learning. After that, the annotation model can generate a natural language annotation which represents the semantics of a given code snippet. Then it combines annotation and original code to represent code vector using the same architecture as the former code search model. The query encoder is also implemented with bidirectional LSTM.

\paragraph{CO3} It also uses multi-task learning to improve the code search model. Inspired by CoaCor~\citep{Yao2019}, besides code annotation, CO3 introduces code generation which has a duality property~\citep{xia2017dual} by nature, with code annotation at the same time. This approach combines the three different code-related tasks, and uses the dual relationship between code generation and code annotation to train the three models simultaneously. It follows CoaCor and also uses bidirectional LSTM for both code encoder and query encoder.


\begin{table*}[h!]
\centering
{\small
\renewcommand\arraystretch{1.2}
\caption{Overall performance of COSEA and baseline methods on StaQC.}
\label{tab:overall_table}
\setlength{\tabcolsep}{0.9mm}{\begin{tabular}{cccc|ccc}
\toprule
 & \multicolumn{3}{c|}{\textbf{StaQC-Python}} & \multicolumn{3}{c}{\textbf{StaQC-SQL}} \\ 
\textbf{Model}& \textbf{Precision@1}& \textbf{MRR}& \textbf{NDCG} &\textbf{Precision@1}& \textbf{MRR}& \textbf{NDCG}\\ 
\hline
CODEnn& 0.547 $\pm$ 0.007& 0.677 $\pm$ 0.003& 0.751 $\pm$ 0.002&0.384 $\pm$ 0.013& 0.538 $\pm$ 0.010& 0.641 $\pm$ 0.008\\ 
CODEnn FT \footnote[1]& 0.584 $\pm$ 0.002& 0.711 $\pm$ 0.001& 0.778 $\pm$ 0.001& 0.422 $\pm$ 0.006& 0.569 $\pm$ 0.007& 0.665 $\pm$ 0.006\\ 
UNIF& 0.608 $\pm$ 0.002& 0.728 $\pm$ 0.002& 0.791 $\pm$ 0.001& 0.377 $\pm$ 0.005& 0.525 $\pm$ 0.003& 0.630 $\pm$ 0.002\\ 
SA& 0.626 $\pm$ 0.001& 0.737 $\pm$ 0.002& 0.796 $\pm$ 0.002 & 0.387 $\pm$ 0.005& 0.539 $\pm$ 0.003& 0.641 $\pm$ 0.003\\ 
CoaCor& --& 0.636& 0.721& --& 0.576& 0.670\\ 
CO3& --& 0.682& 0.756& --& 0.585& 0.679\\ 
\textbf{COSEA}& \textbf{0.657 $\pm$ 0.003}& \textbf{0.764 $\pm$ 0.001}& \textbf{0.819 $\pm$ 0.002}& \textbf{0.445 $\pm$ 0.002}& \textbf{0.587 $\pm$ 0.001 }& \textbf{0.680 $\pm$ 0.001}\\
\bottomrule
\end{tabular}}
}
\end{table*}
\footnotetext[1]{CODEnn FT refers to CODEnn with FastText\citep{bojanowski2017enriching}, through the comparison of CODEnn and CODEnn FT, we can see the importance of pre-trained embeddings.}

\subsection{Hyper Parameters and Settings} \label{sec:hyper}
In the experiments, we set the convolutional hidden space dimension, code word embedding dimension and natural language query word embedding dimension as 400, 200 and 200, respectively. According to the statistical analysis of the StaQC dataset, we set the maximal length of the code sequence to 200 and the maximal length of the natural language query to 20. $\varepsilon$ is fixed to 0.2 for COSEA and all implemented baseline methods except for CoaCor and CO3, because these two methods are hard to implement due to their high complexity and the unavailability of source codes for CO3. We find that a finer tuning on $\varepsilon$ can significantly improve the baseline performance, while there has been no previous studies exploring that.  In CoaCor and CO3, $\varepsilon$ is fixed to $0.05$ as reported by the references. Regarding the training part of the neural networks, we use Adam as the parameter optimizer and the initial learning rate is set to 1e-3. 
For the tokenization method, the original method in work \citep{yao2018staqc} are adopted for codes, and NLTK is used for queries. As pointed out in UNIF \citep{Cambronero2019}, the FastText pre-trained embedding can improve the performance, we adopt FastText as the default method for generating the token embedding for query and code preprocessing.
The whole COSEA method and other baseline methods are implemented using PyTorch and trained on Tesla V100 GPU.

\subsection{Evaluation Metrics} 
In order to evaluate the performance of the model, for each query, we obscure the ground-truth code with 49 randomly selected ones from the entire code sets and obtain the correlation rank of the total 50 codes using the output similarity by the model. Particularly, to avoid information disclosure, we ensure that all 50 candidate codes for each query in the validation and testing process do not appear in the training set. The ranking results will be evaluated using the following metrics:
\paragraph{Precision@$k$}  Precision@$k$ is the probability that the top $k$ codes in the search results contain the ground-truth code, that is
\begin{equation}
    \text{Precision}@k = \frac{1}{|\mathcal{Q}|} \sum_{q\in \mathcal{Q}} \delta(\text{FRank}_q\leq k),
\end{equation}
where $\text{FRank}_q$ is the ranking of the ground-truth code for query $q$ and $\delta$ is an indicative function, which equals to 1 if the input is true and otherwise 0. 

\paragraph{Mean Reciprocal Rank (MRR)} MRR is proposed in work \citep{voorhees1999trec} and it calculates the average value of the reciprocal ranking of the ground-truth code in the ranking results. Thus,
\begin{equation}
    \text{MRR} = \frac{1}{|\mathcal{Q}|} \sum^{|\mathcal{Q}|}_{q=1} \frac{1}{\text{FRank}_{q}}
\end{equation}
\paragraph{Normalized Discounted Cumulative Gain (NDCG)} NDCG defines a revenue score $r(j)$ for the each ranking $j$, and adds all the revenue to obtain a cumulative revenue, and then maps the cumulative revenue to $ [0,1] $ through the normalization function. The formal definition is 
\begin{equation}
    N(n) = Z_{n}\left(\sum^{n}_{j=1} \frac{(2^{r(j)}-1)}{\log(1+j)}\right),
\end{equation}
where $Z_n$ is the normalization function.
In our setting, $n$ equals to 50 and $r(j) = \delta(\text{FRank}_q = 1 \text{ and } j=1)$ for each query $q$. The final NDCG value is the average of all queries' NDCG scores.


\section{Results}\label{results}
We now present our experimental results and answer each of the research questions. All experiments results are listed in the Table \ref{tab:overall_table}. Note that, those results corresponding to CoaCor and CO3 are cited from the original papers since it is hard to reproduce their results due to the high complexity of their methods and unavailability of the source codes. Except CoaCor and CO3, all the other baseline methods are reproduced in our experiments. As aforementioned in Section \ref{sec:hyper}, all these reproduced baseline methods are enhanced with $\varepsilon$ fixed to $0.2$. This can largely improve the performance.
Since previous work only described the proportions for splitting training, validation and testing sets without releasing the already-split datasets to their own experiments, we take the same division proportion and verify the variance of the performance through repeating experiments over random data division. Both the mean performance and corresponding standard deviation for repeated experiments are listed in Table \ref{tab:overall_table}.

\subsection{RQ1}\label{rq1}
\textit{How to capture the semantic information of code more accurately? Is the proposed solution more effective than existing state-of-the-art solutions?}

As detailed in Section \ref{method}, we adopt a convolutional neural network to learn the block representation for code snippets. With the natural ability to capture locality information of code snippets, COSEA's semantic code encoding network represents code in a process similar to the code understanding process of an experienced programmer. And as disscused in \ref{method}, we designed the layer-wise attention mechanism to learn the relationship among code blocks' representations.

From Table \ref{tab:overall_table}, we can see that COSEA shows a significant advantage over other baseline methods on StaQC-Python under all evaluation metrics. And COSEA achieves even better than complex models (CoaCor and CO3). For StaQC-SQL, our model is also much better than baseline models but is slightly better than the best before, since the data quality is relatively poor~\footnote{Random sampling some query and code pairs, there are a considerable number of pairs have poor relevance.}. Based on the discussion above, we can see that using the elaborately designed code semantic encoding network can capture the semantic information of code more accurately. 

In addition, we also verified the effectiveness of the layer-wise attention mechanism. Denote COSEA$-$ATT as the model without layer-wise attention. We examine COSEA$-$ATT on StaQC-Python with the same hyper parameters and settings before. The validation curves during training and the comparison of final MRR and NDCG scores after convergence are shown in Figure~\ref{fig:curve} and Figure~\ref{fig:ablation} respectively. From the two figures above, we can see that the layer-wise attention mechanism lead to better performance in code search.

The evaluation results in Table \ref{tab:overall_table} prove the effectiveness of COSEA. It is worth finding that COSEA can be further enhanced with multi-task learning techniques introduced by CoaCor and CO3\citep{Yao2019,Ye2020}, and it is hopeful to produce more powerful code search model. 

\subsection{RQ2}\label{rq2}
\textit{How to improve the training speed of code search task? Have the proposed solutions actually improved the performance?}

As we discussed in Section \ref{method}, COSEA is based entirely on convolutional neural networks. Compared to previous RNN-based models, COSEA is capable to make all computations parallelized during training and to better exploit the GPU hardware. The optimization also will be easier due to the number of non-linearities is fixed and independent of the input length. 

\begin{figure}[htbp]
 \centering
 \includegraphics[width=0.9\linewidth]{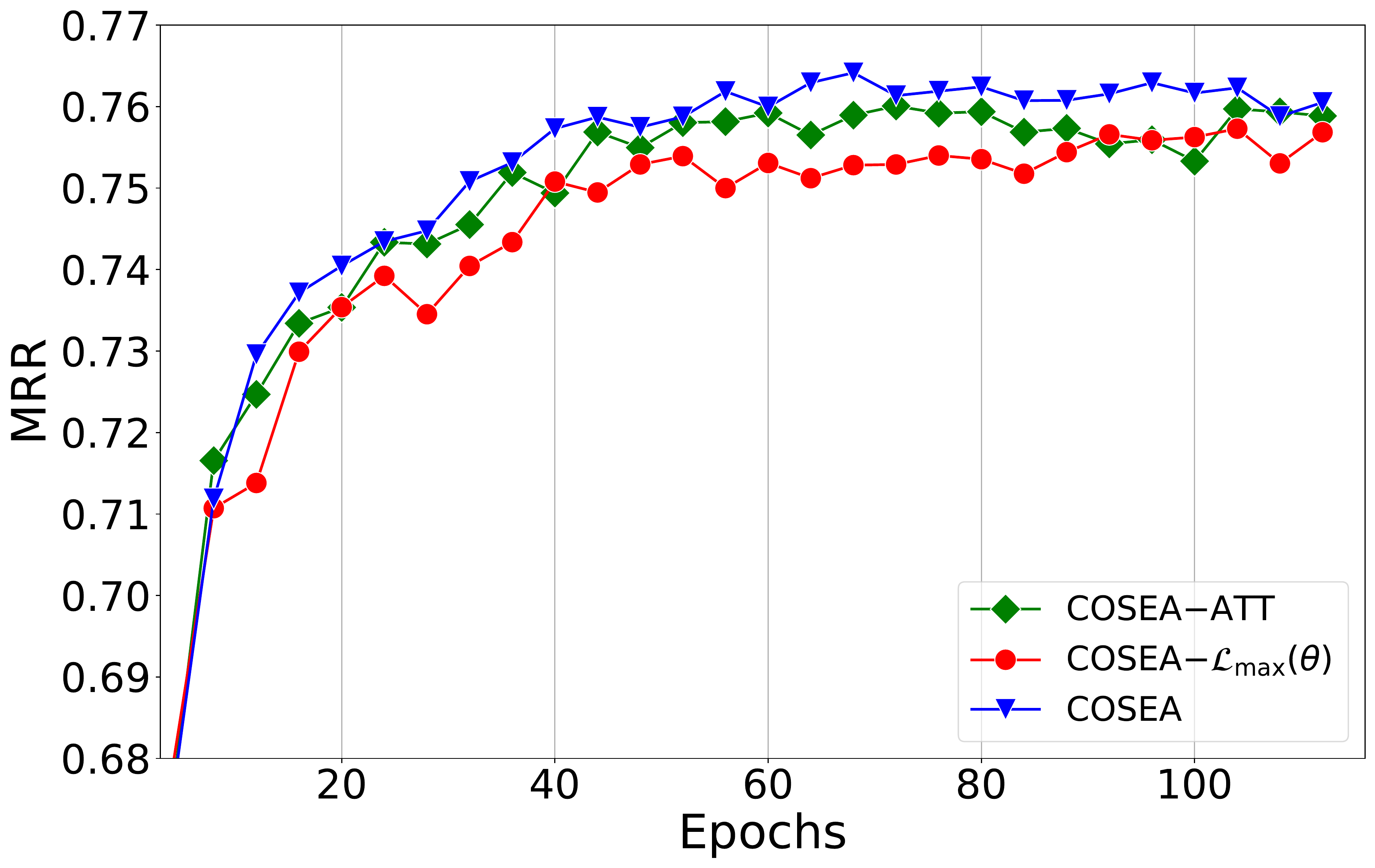}
 \caption{Validation curves of ablation study for COSEA on StaQC-Python. The full version of COSEA show considerable advantages on convergence speed and test MRR score.}\label{fig:curve}
\end{figure}

\begin{figure}[htbp]
 \centering
 \includegraphics[width=0.9\linewidth]{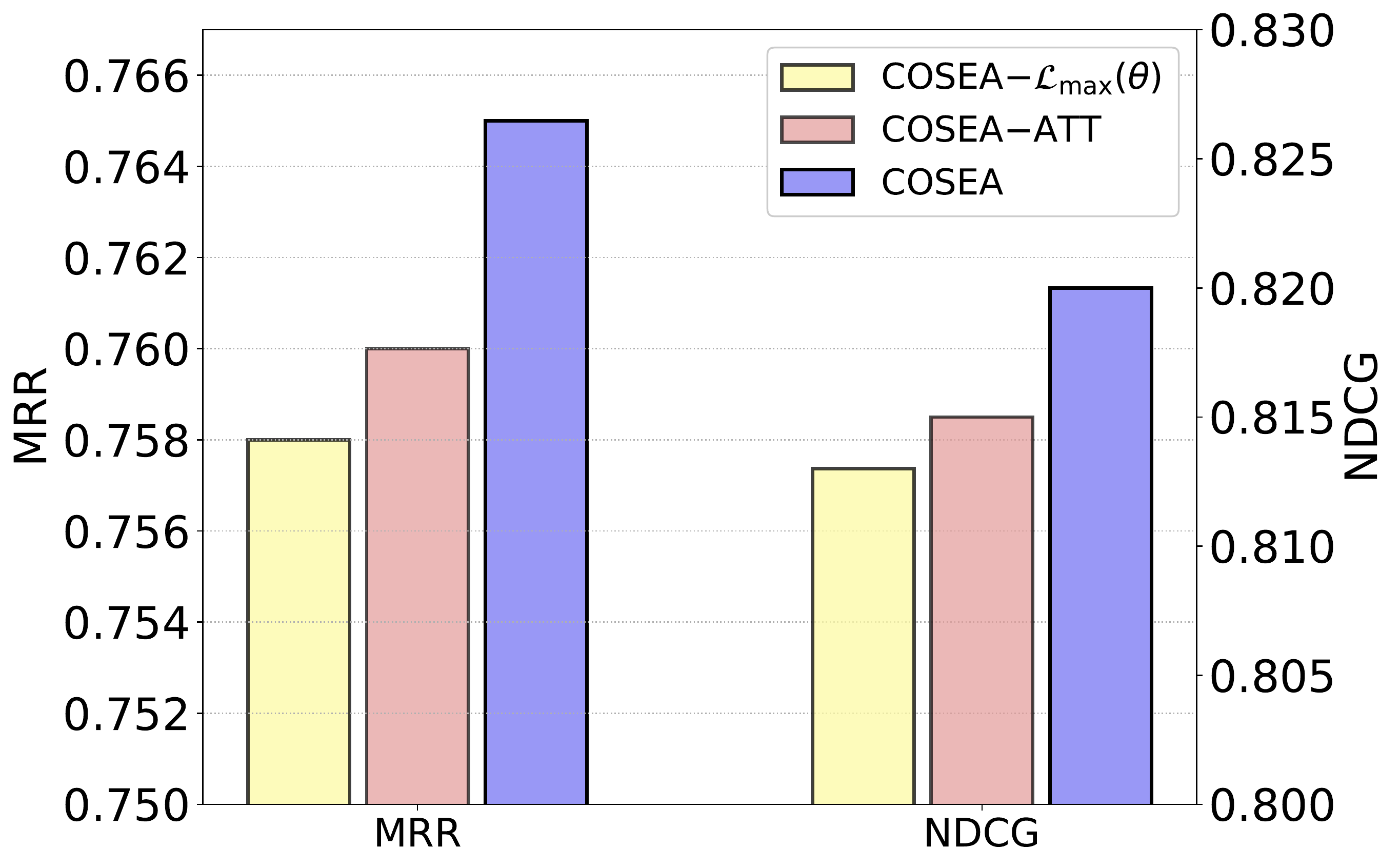}
 \caption{Convergent MRR and NDCG scores comparison of ablation study for COSEA on StaQC-Python.}\label{fig:ablation}
\end{figure}

We further verified the effectiveness of the min-max contrastive learning objective. Denote COSEA$-\mathcal{L}_{\max}(\theta)$ as the model without the min-max contrastive learning objective. We examine this COSEA variant on StaQC-Python with the same hyper parameters and settings. The validation curves during training are shown in Figure~\ref{fig:curve}. From this figure, we can see that the min-max contrastive learning object has demonstrated its considerable advantages in terms of the convergence speed and resulting MRR score. The comparison of final MRR and NDCG scores after convergence on StaQC-Python is shown in Figure~\ref{fig:ablation}. From this figure, we can find that COSEA can significantly outperform this variant. We can conclude that the min-max contrastive learning objective indeed contribute to faster convergence and better performance.

\subsection{RQ3}\label{rq3}
\textit{What does the deep learning model actually learn when dealing with the code search task? Does the knowledge learned by the model make sense?}

\begin{figure}[htbp]
 \centering
  \begin{subfigure}[ht]{1\linewidth}\label{ea} 
    \includegraphics[width=1\textwidth]{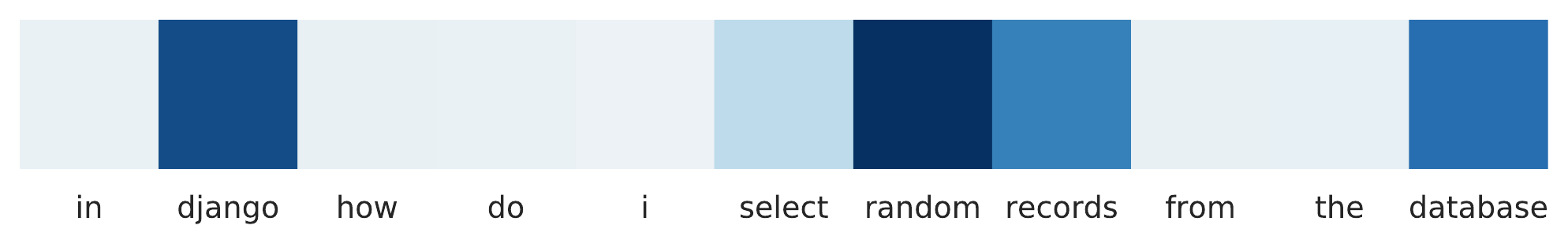}
    \includegraphics[width=1\textwidth]{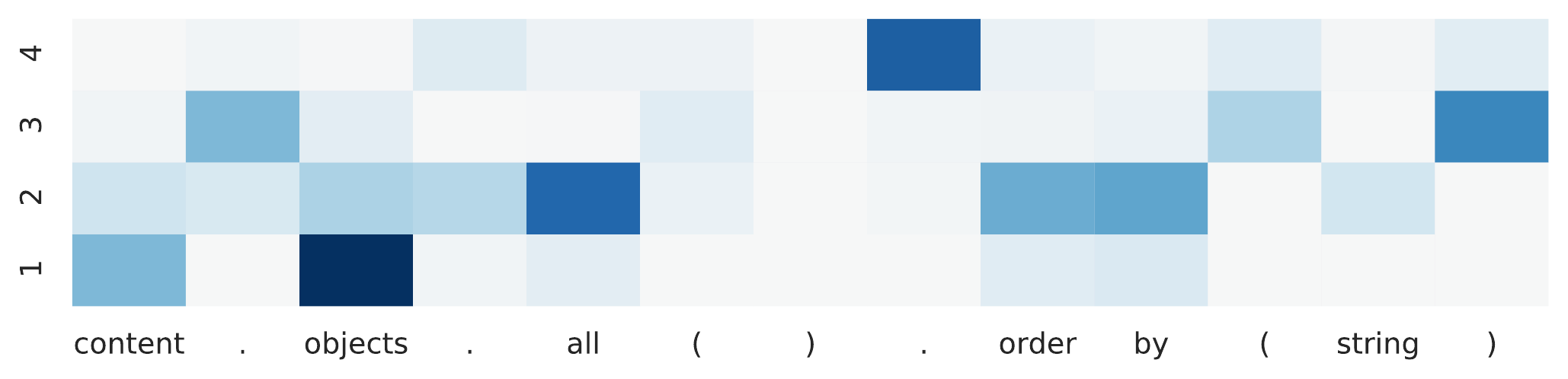}
    \caption{Example from StaQC-Python}
  \end{subfigure}
  \begin{subfigure}[ht]{1\linewidth}\label{eb} 
    \includegraphics[width=1\textwidth]{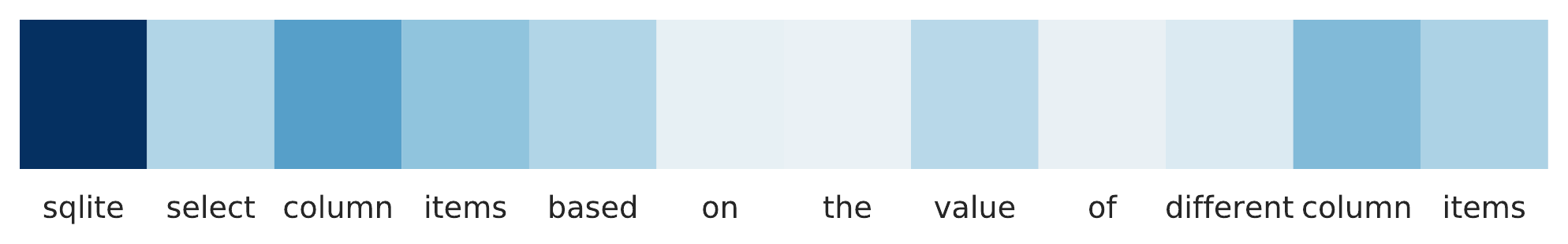}
    \includegraphics[width=1\textwidth]{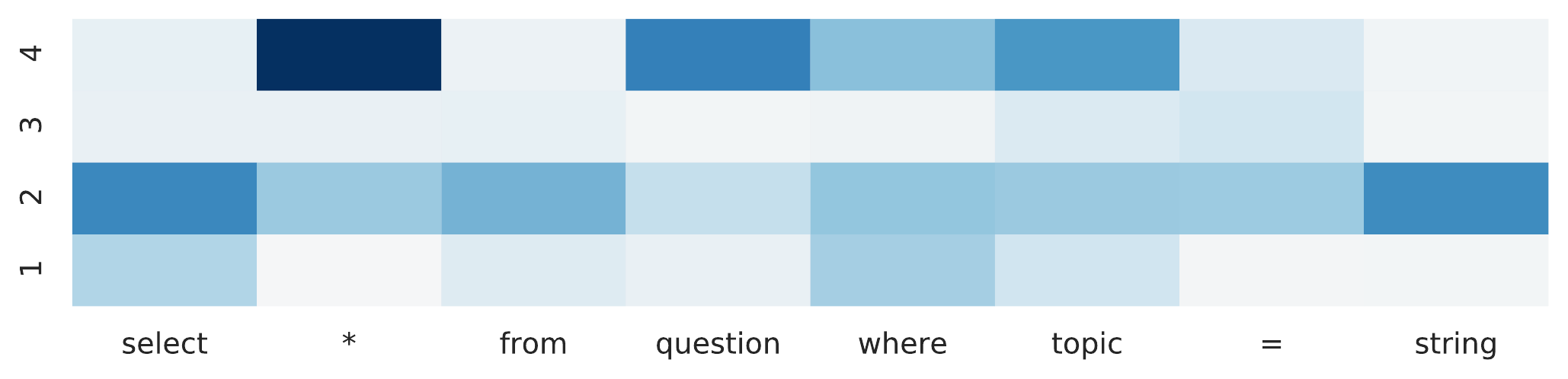}
    \caption{Example from StaQC-SQL}
  \end{subfigure}
 \caption{Heatmap of Query and Code Semantic Representation in attention process of COSEA.}
\label{fig:case_study}%
\end{figure}

Different from the previous work, although we all use the deep learning model, COSEA explicitly contains the layer-wise attention for learning relationship among code blocks and the attention for learning relationship among words of queries. We can check the attention representation of COSEA to qualitatively analyze what COSEA has learned. We transform the attention representation of code and queries into heat map. Two examples are listed in Figure \ref{fig:case_study}, which are seleted from StaQC-Python and StaQC-SQL respectively. And the code snippets are both the first result of each query.

In example \ref{fig:case_study} (a), we can find that COSEA pays more attention on "django", "select", "random", "record" and "database" in this query. For code snippets, COSEA pays more attention to "content.objects.all" and "all().orderby". In example \ref{fig:case_study} (b), we can see that COSEA focuses on "sqlite", "column", "items", "value" in this query, and pays more attention to "select * from", "question where topic", "topic = string" in this code snippets. It's worth noting that the query and code snippets nearly do not share keywords. From the attention heat map and the code search result, we can see that COSEA really has learned how to represent code and query when dealing with the code search task like we designed in the beginning.

\section{Conclusion and Future Work}
In this work, we studied the code search task, where for a natural language query, the related codes should output from a code base. We proposed COSEA, a convolutional model with layer-wise attention, which can fully capture the intrinsic block property of code. To further increase the learning efficiency and performance of the code search task, we introduce a novel min-max contrastive learning objective. Comprehensive experiments on the dataset StaQC-Python and StaQC-SQL elaborate the significant advantage of COSEA over other baseline approaches and state-of-art complex models, as well as the importance of the layer-wise attention and min-max contrastive learning objective. For further study, combining COSEA with the multi-task learning techniques is a promising direction, since CoaCor and CO3 proved multi-task learning can largely improve those RNN-based models. 
In addition, as the BERT model has shown a great advantage in NLP tasks, replacing FastText token embedding with BERT in COSEA implementation also has great potential in code search tasks.





\bibliography{bib/reference}

\begin{thebibliography}{35}
\providecommand{\natexlab}[1]{#1}
\providecommand{\url}[1]{\texttt{#1}}
\expandafter\ifx\csname urlstyle\endcsname\relax
  \providecommand{\doi}[1]{doi: #1}\else
  \providecommand{\doi}{doi: \begingroup \urlstyle{rm}\Url}\fi

\bibitem[Bahdanau et~al.(2014)Bahdanau, Cho, and Bengio]{bahdanau2014neural}
Dzmitry Bahdanau, Kyunghyun Cho, and Yoshua Bengio.
\newblock Neural machine translation by jointly learning to align and
  translate.
\newblock \emph{arXiv preprint arXiv:1409.0473}, 2014.

\bibitem[Bojanowski et~al.(2017)Bojanowski, Grave, Joulin, and
  Mikolov]{bojanowski2017enriching}
Piotr Bojanowski, Edouard Grave, Armand Joulin, and Tomas Mikolov.
\newblock Enriching word vectors with subword information.
\newblock \emph{Transactions of the Association for Computational Linguistics},
  5:\penalty0 135--146, 2017.

\bibitem[Cambronero et~al.(2019)Cambronero, Kim, and Chandra]{Cambronero2019}
Jose Cambronero, Seohyun Kim, and Satish Chandra.
\newblock {When Deep Learning Met Code Search}.
\newblock 2019.

\bibitem[Dauphin et~al.(2017)Dauphin, Fan, Auli, and
  Grangier]{dauphin2017language}
Yann~N Dauphin, Angela Fan, Michael Auli, and David Grangier.
\newblock Language modeling with gated convolutional networks.
\newblock In \emph{Proceedings of the 34th International Conference on Machine
  Learning-Volume 70}, pages 933--941. JMLR. org, 2017.

\bibitem[Devlin et~al.(2018)Devlin, Chang, Lee, and Toutanova]{devlin2018bert}
Jacob Devlin, Ming-Wei Chang, Kenton Lee, and Kristina Toutanova.
\newblock Bert: Pre-training of deep bidirectional transformers for language
  understanding.
\newblock \emph{arXiv preprint arXiv:1810.04805}, 2018.

\bibitem[Efthimiadis(1996)]{efthimiadis1996query}
Efthimis~N Efthimiadis.
\newblock Query expansion.
\newblock \emph{Annual review of information science and technology (ARIST)},
  31:\penalty0 121--87, 1996.

\bibitem[Er et~al.(2016)Er, Zhang, Wang, and Pratama]{er2016attention}
Meng~Joo Er, Yong Zhang, Ning Wang, and Mahardhika Pratama.
\newblock Attention pooling-based convolutional neural network for sentence
  modelling.
\newblock \emph{Information Sciences}, 373:\penalty0 388--403, 2016.

\bibitem[Feng et~al.(2020)Feng, Guo, Tang, Duan, Feng, Gong, Shou, Qin, Liu,
  Jiang, and Zhou]{Feng2020}
Zhangyin Feng, Daya Guo, Duyu Tang, Nan Duan, Xiaocheng Feng, Ming Gong, Linjun
  Shou, Bing Qin, Ting Liu, Daxin Jiang, and Ming Zhou.
\newblock {CodeBERT: A Pre-Trained Model for Programming and Natural
  Languages}.
\newblock 2020.
\newblock URL \url{http://arxiv.org/abs/2002.08155}.

\bibitem[Frome et~al.(2013)Frome, Corrado, Shlens, Bengio, Dean, Ranzato, and
  Mikolov]{frome2013devise}
Andrea Frome, Greg~S Corrado, Jon Shlens, Samy Bengio, Jeff Dean, Marc'Aurelio
  Ranzato, and Tomas Mikolov.
\newblock Devise: A deep visual-semantic embedding model.
\newblock In \emph{Advances in neural information processing systems}, pages
  2121--2129, 2013.

\bibitem[Gehring et~al.(2017)Gehring, Auli, Grangier, Yarats, and
  Dauphin]{gehring2017convolutional}
Jonas Gehring, Michael Auli, David Grangier, Denis Yarats, and Yann~N Dauphin.
\newblock Convolutional sequence to sequence learning.
\newblock In \emph{Proceedings of the 34th International Conference on Machine
  Learning-Volume 70}, pages 1243--1252. JMLR. org, 2017.

\bibitem[Gu et~al.(2018)Gu, Zhang, and Kim]{gu2018deep}
Xiaodong Gu, Hongyu Zhang, and Sunghun Kim.
\newblock Deep code search.
\newblock In \emph{2018 IEEE/ACM 40th International Conference on Software
  Engineering (ICSE)}, pages 933--944. IEEE, 2018.

\bibitem[Haiduc et~al.(2013)Haiduc, Bavota, Marcus, Oliveto, {De Lucia}, and
  Menzies]{Haiduc2013}
Sonia Haiduc, Gabriele Bavota, Andrian Marcus, Rocco Oliveto, Andrea {De
  Lucia}, and Tim Menzies.
\newblock {Automatic query reformulations for text retrieval in software
  engineering}.
\newblock \emph{Proceedings - International Conference on Software
  Engineering}, pages 842--851, 2013.
\newblock ISSN 02705257.
\newblock \doi{10.1109/ICSE.2013.6606630}.

\bibitem[He et~al.(2016)He, Zhang, Ren, and Sun]{he2016deep}
Kaiming He, Xiangyu Zhang, Shaoqing Ren, and Jian Sun.
\newblock Deep residual learning for image recognition.
\newblock In \emph{Proceedings of the IEEE conference on computer vision and
  pattern recognition}, pages 770--778, 2016.

\bibitem[Karpathy and Fei-Fei(2015)]{karpathy2015deep}
Andrej Karpathy and Li~Fei-Fei.
\newblock Deep visual-semantic alignments for generating image descriptions.
\newblock In \emph{Proceedings of the IEEE conference on computer vision and
  pattern recognition}, pages 3128--3137, 2015.

\bibitem[Kim(2017)]{kim2017convolutional}
Phil Kim.
\newblock Convolutional neural network.
\newblock In \emph{MATLAB deep learning}, pages 121--147. Springer, 2017.

\bibitem[Kim(2014)]{kim2014convolutional}
Yoon Kim.
\newblock Convolutional neural networks for sentence classification.
\newblock \emph{arXiv preprint arXiv:1408.5882}, 2014.

\bibitem[Krizhevsky et~al.(2012)Krizhevsky, Sutskever, and
  Hinton]{krizhevsky2012imagenet}
Alex Krizhevsky, Ilya Sutskever, and Geoffrey~E Hinton.
\newblock Imagenet classification with deep convolutional neural networks.
\newblock In \emph{Advances in neural information processing systems}, pages
  1097--1105, 2012.

\bibitem[Le and Mikolov(2014)]{le2014distributed}
Quoc Le and Tomas Mikolov.
\newblock Distributed representations of sentences and documents.
\newblock In \emph{International conference on machine learning}, pages
  1188--1196, 2014.

\bibitem[Liu(2009)]{liu2009learning}
Tie-Yan Liu.
\newblock Learning to rank for information retrieval.
\newblock \emph{Foundations and trends in information retrieval}, 3\penalty0
  (3):\penalty0 225--331, 2009.

\bibitem[Liu et~al.(2019)Liu, Ott, Goyal, Du, Joshi, Chen, Levy, Lewis,
  Zettlemoyer, and Stoyanov]{liu2019roberta}
Yinhan Liu, Myle Ott, Naman Goyal, Jingfei Du, Mandar Joshi, Danqi Chen, Omer
  Levy, Mike Lewis, Luke Zettlemoyer, and Veselin Stoyanov.
\newblock Roberta: A robustly optimized bert pretraining approach.
\newblock \emph{arXiv preprint arXiv:1907.11692}, 2019.

\bibitem[Lu et~al.(2015)Lu, Sun, Wang, Lo, and Duan]{lu2015query}
Meili Lu, Xiaobing Sun, Shaowei Wang, David Lo, and Yucong Duan.
\newblock Query expansion via wordnet for effective code search.
\newblock In \emph{2015 IEEE 22nd International Conference on Software
  Analysis, Evolution, and Reengineering (SANER)}, pages 545--549. IEEE, 2015.

\bibitem[Lv et~al.(2015)Lv, Zhang, Lou, Wang, Zhang, and Zhao]{lv2015codehow}
Fei Lv, Hongyu Zhang, Jian-guang Lou, Shaowei Wang, Dongmei Zhang, and Jianjun
  Zhao.
\newblock Codehow: Effective code search based on api understanding and
  extended boolean model (e).
\newblock In \emph{2015 30th IEEE/ACM International Conference on Automated
  Software Engineering (ASE)}, pages 260--270. IEEE, 2015.

\bibitem[McMillan et~al.(2011)McMillan, Grechanik, Poshyvanyk, Xie, and
  Fu]{mcmillan2011portfolio}
Collin McMillan, Mark Grechanik, Denys Poshyvanyk, Qing Xie, and Chen Fu.
\newblock Portfolio: finding relevant functions and their usage.
\newblock In \emph{Proceedings of the 33rd International Conference on Software
  Engineering}, pages 111--120, 2011.

\bibitem[Mikolov et~al.(2013)Mikolov, Chen, Corrado, and Dean]{Mikolov2013}
Tomas Mikolov, Kai Chen, Greg Corrado, and Jeffrey Dean.
\newblock {Efficient estimation of word representations in vector space}.
\newblock \emph{1st International Conference on Learning Representations, ICLR
  2013 - Workshop Track Proceedings}, pages 1--12, 2013.

\bibitem[Miller(1995)]{miller1995wordnet}
George~A Miller.
\newblock Wordnet: a lexical database for english.
\newblock \emph{Communications of the ACM}, 38\penalty0 (11):\penalty0 39--41,
  1995.

\bibitem[Sachdev et~al.(2018)Sachdev, Li, Luan, Kim, Sen, and
  Chandra]{Sachdev2018}
S.~Saksham Sachdev, H.~Hongyu Li, S.~Sifei Luan, S.~Seohyun Kim, K.~Koushik
  Sen, and S.~Satish Chandra.
\newblock {Retrieval on source code: A neural code search}.
\newblock \emph{MAPL 2018 - Proceedings of the 2nd ACM SIGPLAN International
  Workshop on Machine Learning and Programming Languages, co-located with PLDI
  2018}, pages 31--41, 2018.

\bibitem[Sutskever et~al.(2014)Sutskever, Vinyals, and
  Le]{sutskever2014sequence}
Ilya Sutskever, Oriol Vinyals, and Quoc~V Le.
\newblock Sequence to sequence learning with neural networks.
\newblock In \emph{Advances in neural information processing systems}, pages
  3104--3112, 2014.

\bibitem[Turian et~al.(2010)Turian, Ratinov, and Bengio]{turian2010word}
Joseph Turian, Lev Ratinov, and Yoshua Bengio.
\newblock Word representations: a simple and general method for semi-supervised
  learning.
\newblock In \emph{Proceedings of the 48th annual meeting of the association
  for computational linguistics}, pages 384--394, 2010.

\bibitem[Vaswani et~al.(2017)Vaswani, Shazeer, Parmar, Uszkoreit, Jones, Gomez,
  Kaiser, and Polosukhin]{vaswani2017attention}
Ashish Vaswani, Noam Shazeer, Niki Parmar, Jakob Uszkoreit, Llion Jones,
  Aidan~N Gomez, {\L}ukasz Kaiser, and Illia Polosukhin.
\newblock Attention is all you need.
\newblock In \emph{Advances in neural information processing systems}, pages
  5998--6008, 2017.

\bibitem[Voorhees et~al.(1999)]{voorhees1999trec}
Ellen~M Voorhees et~al.
\newblock The trec-8 question answering track report.
\newblock In \emph{Trec}, volume~99, pages 77--82, 1999.

\bibitem[Xia et~al.(2017)Xia, Qin, Chen, Bian, Yu, and Liu]{xia2017dual}
Yingce Xia, Tao Qin, Wei Chen, Jiang Bian, Nenghai Yu, and Tie-Yan Liu.
\newblock Dual supervised learning.
\newblock In \emph{Proceedings of the 34th International Conference on Machine
  Learning-Volume 70}, pages 3789--3798. JMLR. org, 2017.

\bibitem[Xu et~al.(2015)Xu, Xiong, Chen, and Corso]{Xu2015}
Ran Xu, Caiming Xiong, Wei Chen, and Jason~J. Corso.
\newblock {Jointly modeling deep video and compositional text to bridge vision
  and language in a unified framework}.
\newblock \emph{Proceedings of the National Conference on Artificial
  Intelligence}, 3:\penalty0 2346--2352, 2015.

\bibitem[Yao et~al.(2018)Yao, Weld, Chen, and Sun]{yao2018staqc}
Ziyu Yao, Daniel~S Weld, Wei-Peng Chen, and Huan Sun.
\newblock Staqc: A systematically mined question-code dataset from stack
  overflow.
\newblock In \emph{Proceedings of the 2018 World Wide Web Conference}, pages
  1693--1703, 2018.

\bibitem[Yao et~al.(2019)Yao, Peddamail, and Sun]{Yao2019}
Ziyu Yao, Jayavardhan~Reddy Peddamail, and Huan Sun.
\newblock {COACOR: Code annotation for code retrieval with reinforcement
  learning}.
\newblock \emph{The Web Conference 2019 - Proceedings of the World Wide Web
  Conference, WWW 2019}, pages 2203--2214, 2019.
\newblock \doi{10.1145/3308558.3313632}.

\bibitem[Ye et~al.(2020)Ye, Xie, Zhang, Hu, Wang, and Zhang]{Ye2020}
Wei Ye, Rui Xie, Jinglei Zhang, Tianxiang Hu, Xiaoyin Wang, and Shikun Zhang.
\newblock {Leveraging Code Generation to Improve Code Retrieval and
  Summarization via Dual Learning}.
\newblock 2020.
\newblock \doi{10.1145/3366423.3380295}.
\newblock URL
  \url{http://arxiv.org/abs/2002.10198{\%}0Ahttp://dx.doi.org/10.1145/3366423.3380295}.

\end{thebibliography}
\vspace{12pt}
\end{document}